\documentclass[aps,prl,preprint,groupedaddress]{revtex4-1}
\pdfoutput=1
\usepackage{graphicx}
\usepackage{epsfig}
\usepackage{amsmath,amssymb,hyperref,enumerate}
\usepackage{slashed}
\usepackage{amsfonts}
\usepackage{url}
\usepackage{color}
\usepackage{bm}
\usepackage{subfig}
\hypersetup{colorlinks,citecolor= black,linkcolor= black}
\definecolor{nicered}{rgb}{0.7,0.1,0.1}
\definecolor{nicegreen}{rgb}{0.1,0.5,0.1}

\usepackage{setspace}

\allowdisplaybreaks

\def\({\left(}
\def\){\right)}
\def\[{\left[}
\def\]{\right]}

\usepackage[margin=1.in]{geometry}

\begin{document}
\preprint{UMD-PP-017-29}

\title{Constraints on Mirror Models of Dark Matter from Observable Neutron-Mirror Neutron Oscillation }

\author{Rabindra N. Mohapatra$^a$}\author{ and Shmuel Nussinov$^b$}
\affiliation{$^a$Maryland Center for Fundamental Physics and Department of Physics,
University of Maryland, College Park, Maryland 20742, USA}
\affiliation{$^b$ School of Physics and Astronomy,  Tel Aviv University,  Tel Aviv 69978, Israel}

\begin{abstract}  The process of neutron-mirror neutron oscillation, motivated by symmetric mirror dark matter models, is governed by two parameters: $n-n'$ mixing parameter $\delta$ and $n-n'$ mass splitting $\Delta$. For neutron mirror neutron oscillation to be observable, the splitting between their masses $\Delta$ must be small and current experiments lead to $\delta \leq 2\times 10^{-27}$ GeV and 
 $\Delta \leq 10^{-24}$ GeV. We show that in mirror universe models where this process is observable, this small mass splitting constrains the way that one must implement asymmetric inflation to satisfy the limits of Big Bang Nucleosynthesis on the number of effective light degrees of freedom. In particular we find that if asymmetric inflation is implemented by inflaton decay to color or electroweak charged particles, the oscillation is unobservable. Also if one uses SM singlet fields for this purpose, they must be weakly coupled to the SM fields. \end{abstract}
\maketitle
\section{1. Introduction}\label{sec:1}
The possibility that there may be a mirror sector of the standard model (called MSM) with identical particle content and gauge symmetry as the standard model (called SM)~\cite{mirror1} has received a great deal of attention for the reason that the dark matter of the universe may be the lightest baryon (or atom) of the mirror sector. These models assume the existence of $Z_2$ symmetry between the two sectors which keeps the same number of parameters even though the number of particles is doubled. Depending on the status of the $Z_2$ symmetry there arise two possibilities: (i) The mirror symmetry is unbroken so that particle masses in the MSM are the same as in the SM \cite{mirror2}and (ii) The symmetry is spontaneously broken~\cite{mirror3} and mirror sector particles to have similar masses compared to the SM.  An important problem for both classes of models is the large number of light particle degrees of freedom at the epoch of big bang nucleosynthesis $(\nu_a, \gamma; e, \nu^\prime_a, \gamma')$ and possibly $e^\prime$ making $N_{eff}=21.5$ in clear contradiction to the current CMB bound from Planck and WMAP of $N_{eff} \leq 11.5$. A solution to this problem of mirror models was suggested in~\cite{asym} where it was proposed that inflation in mirror models is different from that in conventional cosmology. The inflaton is assumed to be a mirror odd scalar field that couples to particles of both sectors in such a way that after inflation is completed, the reheat temperature of the mirror sector becomes less than that of the SM sector i.e. $T^\prime_{reheat} \simeq  \frac{1}{2} T_{reheat} $, so that the contribution of the light mirror particles to energy density (which goes like $\sim T^4$) at the BBN epoch is highly suppressed. An important implication of this proposal, called asymmetric inflation, is that at some high scale, there must be spontaneous breaking of the $Z_2$ mirror symmetry. This breaking should have some effect on low energy physics of the model. This is what we investigate in this paper in the context of the symmetric mirror model.

In the symmetric mirror models, the neutron ($n$) and the mirror neutron ($n^\prime$)  have same mass in the tree approximation. It has been proposed that if there is a small mixing, $\delta$, between $n$ and $n^\prime$ arising from some higher order induced dimension six operators of the form 
$uddu'd'd'$, there can be oscillations between neutron and the mirror neutron~\cite{BB}. Observation of this process would have dramatic implications for particle physics and cosmology. This process has been looked for in experiments~\cite{expt} and currently there is a lower bound on the transition time for $n\leftrightarrow n'$ of  448 sec. which translates to an upper limit on the mixing parameter $\delta \leq 2\times 10^{-27}$ GeV. Further experiments are being planned~\cite{yuri} to search for this process.

For $n\leftrightarrow n'$ oscillation to be observable, a high degree of degeneracy between $n$ and $n'$ is essential. This is similar to the case of neutron-anti-neutron oscillation~\cite{nnbar} where this degeneracy is guaranteed at the particle physics level by CPT invariance. On the other hand, in the case of $n\leftrightarrow n'$ oscillation, the only symmetry that guarantees their mass equality is a $Z_2$ symmetry which is necessarily broken by the requirement of asymmetric inflation. In viable mirror models where asymmetric inflation is enforced by choice of fields, one would expect some degree of mass splitting between the $n$ and $n'$ masses.  In this paper, we show that unless one is careful in choosing the fields required to implement asymmetric inflation, the $n-n'$ mass splitting can exceed the  value required for having observable $n-n'$ oscillation.

It is known that both symmetric and asymmetric mirror models can have viable dark matter candidates~\cite{foot,bere,MNT}. We comment on the implications of observable $n-n'$ oscillation on the properties of dark matter in the former case. 

This paper is organized as follows: in sec. 2, we discuss the upper limit on the splitting of $n,n'$ for the oscillation to be observable; in sec. 3, we review the field theoretic model that implements asymmetric inflation in the mirror model; in sec. 4, we discuss the implications of the asymmetric inflation model for the magnitude of $n-n'$ mass splitting. In sec. 5, we comment on some aspects of the dark matter physics in this model.

\section{2. Required level of $n-n'$ mass degeneracy for observable $n-n'$ oscillation} 
To find the maximal value of the splitting between $m_n$ and $m_{n'}$ which allows observable $n-n'$  oscillation, we ignore spin and write the evolution equation of the $n$ and $n'$ system:
\begin{eqnarray}
\frac{d}{dt}\left(\begin{array}{c} n \\ {n'}\end{array}\right)~=~\left(\begin{array}{cc} m & \delta \\ \delta & m'\end{array}\right)\left(\begin{array}{c} n \\ {n'}\end{array}\right).
\end{eqnarray}
Starting initially with neutrons, the probability that an mirror neutron will appear after a time of $t$ is given by:
\begin{eqnarray}
P_{n-{n'}}~=~\frac{4\delta^2}{\Delta^2+4\delta^2}{\rm sin}^2 \frac{\sqrt{\Delta^2+4\delta^2} ~t/2}{\hbar}
\end{eqnarray}
where $\Delta =m'-m$. This difference could arise from higher order quantum corrections or a difference between magnetic fields ($B$ and $B'$). Here we focus on the first source. Note that assuming  a typical neutron transit time $\sim 1$ sec., we would expect that for $n-n'$ oscillation to be observable, we must have $\Delta \leq 10^{-24}$ GeV. If the transit time is shorter, requiring a more intense neutron beam, the bound would be proportionately weaker.
The neutrons mass is given by $m_n =M_0 +2m_d+m_u$ with $M_0$ being a QCD generated mass and  $m_i= y_i.v$ are the bare quark masses generated by coupling to the Higgs VeV $v$. Since the quark masses contribute only ~ 1\% to the $m_n$ requiring $\Delta< 10 ^{-24}$ GeV implies  that $\delta y/y\equiv (y-y')/y \leq 10^{-22}$  GeV.

We now discuss the impact of $Z_2$ mirror parity breaking for asymmetric inflation and its impact on $\Delta$. We envision a general set-up~\cite{asym}, where we have a $Z_2$ odd gauge singlet field $\eta$ which acts as the inflaton and  $X, X'$ be the two mirror partner complex scalar fields. We assume the following inflaton field potential:
\begin{eqnarray}
V(\eta, X, X')~=~m^2_\eta \eta^2+\lambda_\eta \eta^4+\mu\eta(X^\dagger X-{X'}^{\dagger} X')+\lambda\eta^2(X^\dagger X+{X'}^{\dagger} X')
\end{eqnarray}
 Different inflation pictures arise by modifying the $m^2_\eta \eta^2+\lambda_\eta \eta^4$ part, the details of which are not crucial to our conclusion. The part which leads to asymmetric inflation comes from the last two terms in Eq. (3). To recap the argument of ~\cite{asym}, we note that after $\eta$ field acquires a vev, the inflaton field decays asymmetrically to the $X$ and $X'$ fields with the ratio of the decay widths given by 
\begin{eqnarray}
\frac{\Gamma_{\eta\to XX}}{\Gamma_{\eta\to X'X'}}=\left(\frac{\mu+\lambda<\eta>}{\mu-\lambda<\eta>}\right)^2
\end{eqnarray}
Noticing that reheat temperature is $T_R\simeq \sqrt{\Gamma_\eta M_{Pl}}$, we find that $T'_R < T_R$ if all the parameters are positive. By appropriately choosing the parameters, we can make $T'_R/T_R\simeq 1/2$ as required in order to satisfy BBN constraints. We also note that the required symmetry breaking potential in Eq.(3) leads to different masses for $X,X'$ and to the relation (assuming the bare masses for $X,X'$ to be small):
\begin{eqnarray}M_{X}/M_{X'}\propto T_R/{T'}_R
\end{eqnarray}

For a detailed discussion of reheating in such scenarios, see ~\cite{cui}.

The $X$ and $X'$ must couple to the known SM fields ( and to their mirrors , respectively) in order to generate the latter with the required different temperatures , starting the usual Hubble expansion in each sector. The mass difference of the $X,X'$ fields causes the mirror symmetry breaking manifesting in the different $M_X$ and $M_{X'}$ to trickle down to lower energies and , in particular to generate different Yukawa couplings $y_i$ for the quarks. 

\section{4. Implications for $n-n'$ mass splitting}

In this section, we give several examples for the fields $X,X'$ and note the implications for $\delta y/y\equiv (y-y')/y$ for each case. The correction to $\delta y$ arise from two loop self energy corrections to the Yukawa couplings of quarks of the type shown in Fig.1:

\begin{figure}[h!]
	\centering
\includegraphics[scale=0.3]{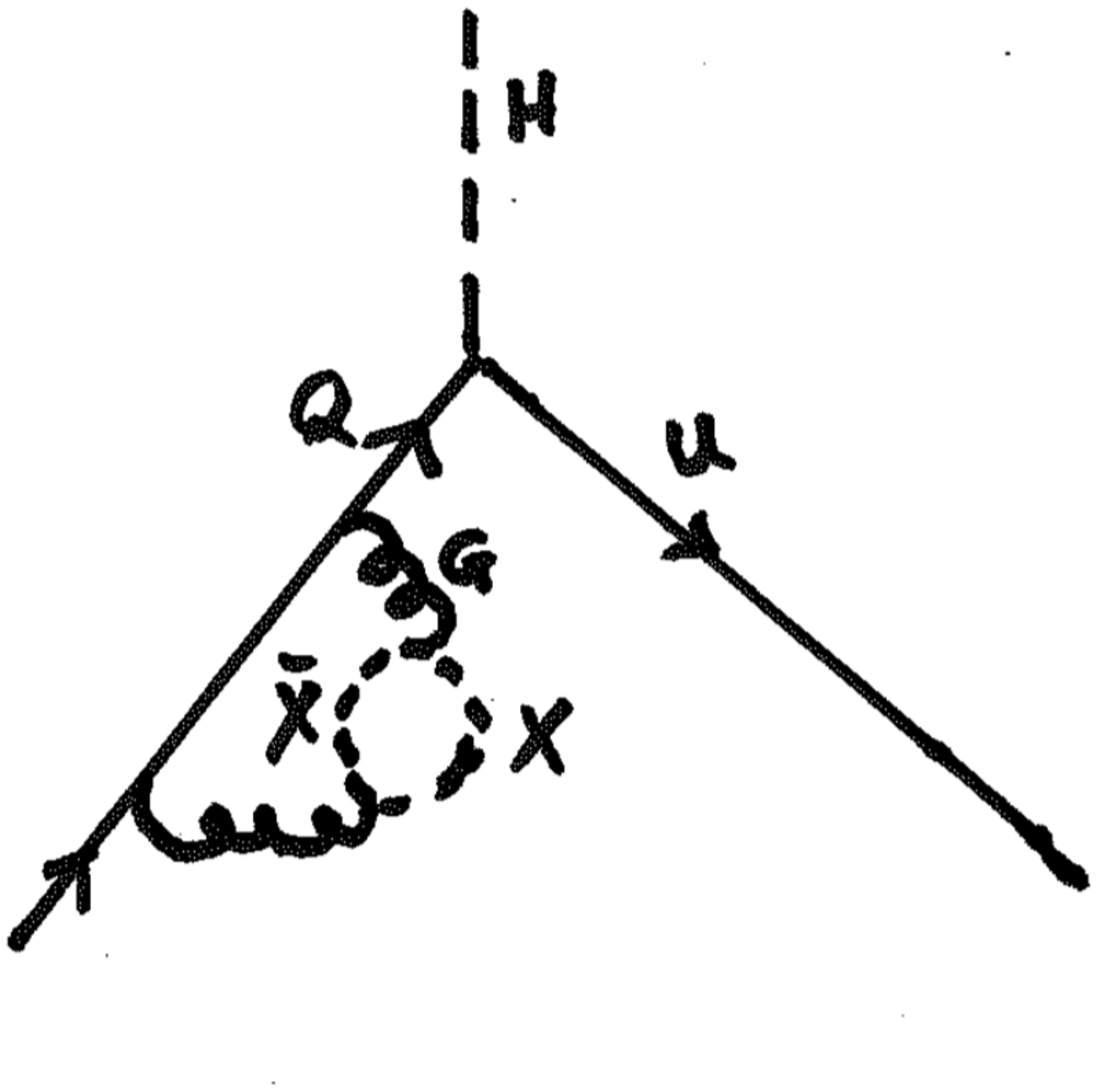}
\caption{Typical two loop diagram that contributes mass difference between $n$ and $n'$. This graph corresponds to the case where $X$ is a colored particle.}
\label{fig1}
\end{figure}

\begin{itemize}

\item If $X$ and $X'$ are color non-singlet fields, then they will have color gauge couplings which will induce different wave function renormalization for the quarks in the Yukawa couplings which through RGE running lead to an estimate $\frac{\delta y}{y}\simeq \frac{\alpha^2_c}{16\pi^2}\ell n\frac{M^2_X}{{M}^2_{X'}}=\frac{\alpha^2_c}{16\pi^2}\ell n\frac{T^2_R}{{T'}^2_{R}}$ (where $\alpha_c$ is the QCD fine structure constant). This splitting is of order $\sim 10^{-4}$, which is clearly much larger than what is allowed in order to make $n-n'$ oscillation observable. Therefore if $n-n'$ oscillation is to be observable, one cannot allow $X,X'$ to have color.

\item Instead if $X,X'$ were color singlet but  $SU(2)_L$ and $SU(2)'_L$ non-singlets, we replace the $\alpha_c$ above by $\alpha_W$ yielding $\frac{\delta y}{y}\sim 10^{-6}$, which is also much larger than $\Delta$ estimated above.

\item This difficulty can be ameliorated if  $X,X'$ are SM and MSM singlets e.g.  a singlet scalar,$S$ and $S'$. In this case the scalar must couple to SM fields to generate the familiar Hubble expansion with SM matter fields. This implies that $S$ couples to the Higgs fields (of SM and MSM) via the coupling $\mu_S (SH^\dagger H+S'{H'}^\dagger H')$. In this case, we have the $\frac{\delta y}{y} \sim \frac{y^2_u\mu^2_S}{(16\pi^2)^2 M^2_S}\ell n\frac{T^2_R}{{T'}^2_{R}}$ and we can bring down this value to the acceptable level for observability of $n-n'$ oscillation by tuning $\mu_S$ somewhat. Taking $y_u\sim 10^{-5}$ and $\mu_S/M_S\sim 10^{-4}$, we get the desired small splitting $\Delta$. The first factor $y_u$ corresponds to the first generation quark Yukawa couplings since $n$ and $n'$ involve only the first generation quarks (Fig 2).

\item The singlet fields could also be singlet fermions in which case the the inflation coupling to them would contain nonrenormalizable terms i.e. ${\cal L}_\eta \sim \eta (NN-N'N')+\frac{\eta^2}{\Lambda}(NN+N'N')$ and all the above considerations go through.

\item In discussing all these cases we have assumed that the bare mass contribution to $X.X'$ masses is small compared to the $<\eta>$ contribution. However, if we assume that the bare masses are much larger than $<\eta>$, the log in the splitting corrections simplifies to $\frac{M^2_X-M^2_{X'}}{M^2_X}$ and for $\delta y/y$ to be less than $10^{-22}$, we must have $\frac{M^2_X-M^2_{X'}}{M^2_X}\leq 10^{-18}$ (in the colored $X,X'$ and slightly weaker bound in the  SM non-singlet case.). This requires that $<\eta>/M_X \leq 10^{-18}$ which is a very high degree of fine tuning. 

\end{itemize}
\begin{figure}[h!]
	\centering
\includegraphics[scale=0.3]{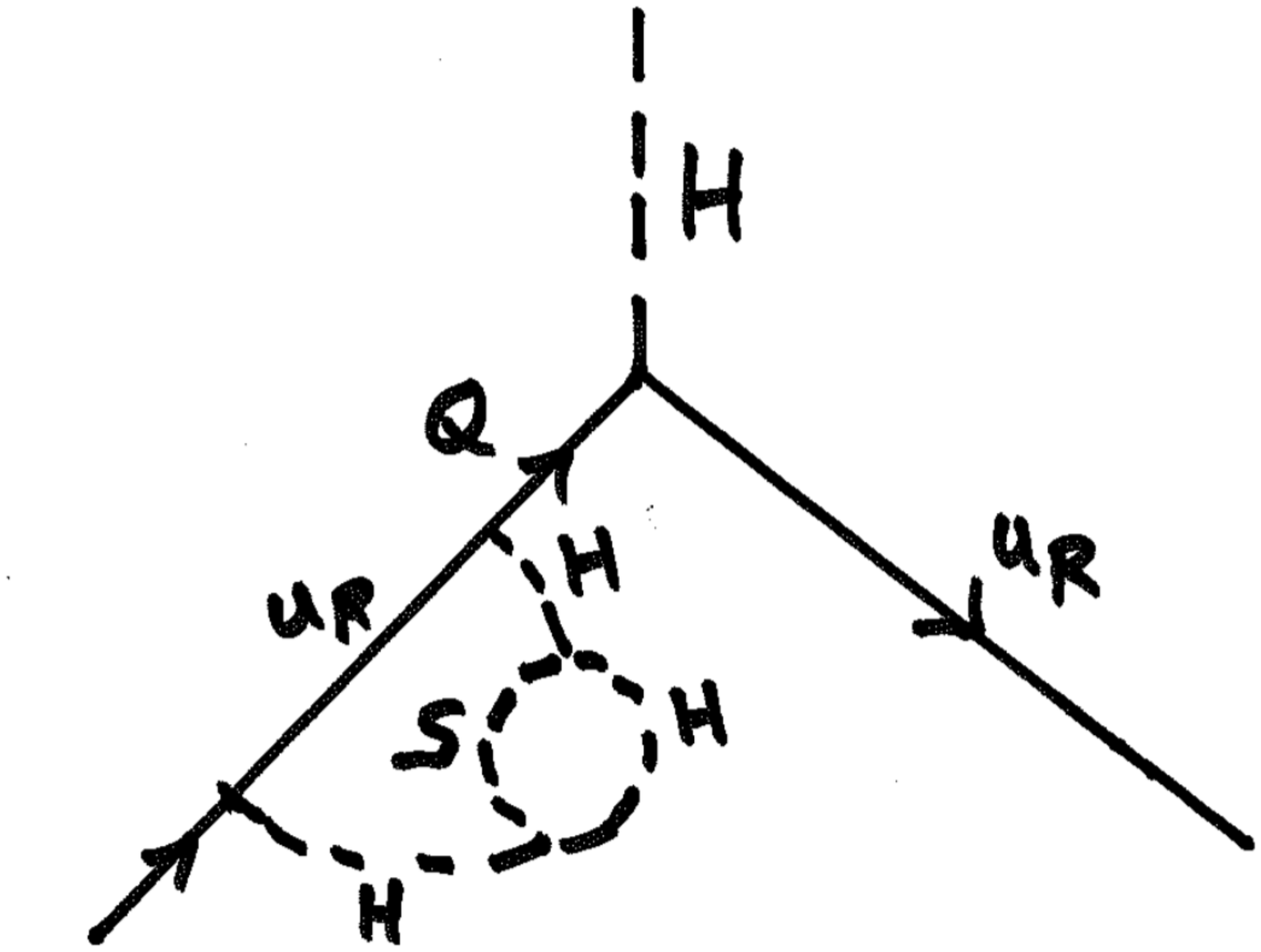}
\caption{Typical two loop diagram that contributes mass difference between $n$ and $n'$. This graph corresponds to the case where $X$ is a SM singlet scalar field.}
\label{fig2}
\end{figure}

From the above discussion, we see that the structure of the inflation sector of the mirror model must be very restrictive i.e. the inflaton field must couple to only to  SM and MSM singlets. Furthermore the coupling of the singlet fields to SM and MSM fields must be very weak. 

\section{5. Comments on dark matter properties in symmetric vs asymmetric mirror models}
Observable $n-n'$ oscillation also has implications for the nature of dark matter as we see now. This is due to the fact that dark matter has self interactions and there are limits on the DM-DM cross section from bullet cluster observations~\cite{bullet}.

In mirror scenarios with appreciable mirror symmetry breaking (e.g. the asymmetric mirror model), the fermion spectrum in the SM and MSM sector need not be same and therefore the lightest mirror nucleon can be $n'$  and serve as the DM particle.The cross-section for low energy $n'-n'$ scattering is
\begin{eqnarray}
 \sigma(n'n') \sim \sigma( nn) = 4\pi a^2 \sim {\rm Barn}=  10 ^{-24} {\rm cm}^2.  
 \end{eqnarray}
 This is an optimal value close to what may be required to solve the CDM cusp problem~\cite{SS} and yet consistent with the Bullet cluster  upper bound.

 This is not the case in models with the very precise mirror symmetry  required for the observability of  n-n' oscillations. To have a kinematically allowed $p'\to n' +\nu' +{e'}^{+}$ decay so as to ensure that $n'$ is the dark matter,  the masses of some mirror particles should be shifted relative to the masses of their SM counter-parts by several MeV- exceeding by twenty orders of magnitude the allowed $n'-n$ splitting $\Delta$.

 The dark matter should then consist of dark atoms i.e. $H'$ and $He'$ - mirror Hydrogen and mirror Helium and it was pointed out that the latter may be more abundant relative to the former than in our ordinary baryonic matter~\cite{bere}. However in either case the low energy DM-DM elastic scattering is fixed by the atomic sizes to be ~$\sim {\rm Angstrom}^2 = 10 ^{-16}$ cm$^2$ which will exceed the upper bound  by 7-8 orders of magnitude. Due to the large impact parameter in the atomic collisions we do not have - as in the case of $n-n $ (or $n'-n'$) scattering -only S wave contribution and the differential elastic cross-section is non-isotropic and forward peaked. This reduces the effective transport cross-section which as noted in~\cite{MNT}  is  the one relevant here. At the relatively high O( keV) collision energies which are $\sim 100$ times the Ionization threshold), there can also be many atomic excitations and ionization processes and theoretical estimates of the total stopping power of the gas are difficult to make. Actual data on atomic beam scattering~\cite{Phelps} suggest about 1/{100}  reduction of the effective cross-sections from the naive Angstrom $^2$ estimate leaving us with values  which are still considerably higher than the upper bound from bullet cluster observation.

 This would suggest that to have a viable model we need that - unlike for ordinary galaxies and galaxies clusters most of the mirror matter should reside in collisonless stars and not in gas- a rather non-trivial constraint when added to other requirements which DM has to satisfy.

In summary, we have pointed out that for $n-n'$ to be observable, one must have a specific structure for the inflation reheat sector of the Lagrangian in mirror models- the inflation reheating must proceed via fields that are standard model  and mirror standard model gauge singlets.
We also make some general comments on the dark matter properties in symmetric vs asymmetric mirror picture, which has bearing on the issue of $n-n'$ oscillation to be observable.

\section*{Acknowledgement}  R.N.M. was supported by the US National Science Foundation under Grant No. PHY1620074.

\end{document}